# Ab initio insights into the interaction mechanisms between $H_2$, $H_2O$, and $O_2$ molecules with diamond surfaces

Nam V. Tran and M.C. Righi*

Department of Physics and Astronomy, University of Bologna, 40127, Bologna, Italy

**Abstract**

Diamond displays outstanding chemical, physical, and tribological properties, making it attractive for numerous applications ranging from biomedicine to tribology. However, the reaction of the materials with molecules present in the air, such as oxygen, hydrogen, and water, could significantly change the electronic and tribological properties of the films. In this study, we performed several density functional theory calculations to construct a database for the adsorption energies and dissociation barriers of these molecules on the most relevant diamond surfaces, including C(111), C(001), and C(110). The adsorption configurations, reaction paths, activation energies, and their influence on the structure of diamond surfaces are discussed. The results indicate that there is a strong correlation between adsorption energy and surface energy. Moreover, we found that the dissociation processes of oxygen molecules on these diamond surfaces can significantly alter the surface morphology and may affect the tribological properties of diamond films. These findings can help to advance the development and optimization of devices and antiwear coatings based on diamond.

# 1. Introduction

Diamond and diamond-like carbon (DLC) are widely studied as coating materials in a number of fields ranging from tribology [1, 2] to biomedicine [3, 4] due to their exceptional chemical, physical, mechanical, biomedical, and tribological properties. In tribological applications, the carbon-based films can provide an extremely low wear rate and coefficients of friction (COF) without using any chemical additives or lubricants, making the carbon coatings advantageous for environmental protection [5-8]. Thanks to these outstanding properties, the materials are now widely utilized in numerous industrial applications such as engine components [3, 4], artificial joints [1, 2], sensors [9], and more recently as structural materials for nano-and microelectromechanical systems (NEMS/MEMS) [10-12].

However, the widespread use of carbon-based films has been deterred due to the difficulty in maintaining and controlling the electronic and tribological properties of the films under different working environments. The reaction of the materials with molecules in the air, such as oxygen, hydrogen, and water, significantly change the electronic and tribological properties of the films [13, 14]. A large number of experimental and theoretical studies have been devoted to investigating the influence of adsorbates such as hydrogen and hydroxyl groups on the electronic and mechanical behaviors of diamond-like materials [15-26]. For example, previous studies showed that the reaction of diamond with water molecules in the air could not only alter several bulk properties of the material, such as its conductivity and dielectric permittivity, but also significantly affect the tribological properties of different diamond surfaces [25-27]. Indeed, hydrophilic hydroxyl groups chemisorbed from the reaction of water with the films are considered to play a crucial role in reducing friction [1, 16, 28-33]. Similarly, the tribological properties and surface reactivity of diamond-like carbon (DLC) films are highly dependent on the H contents in the films [34, 35]. It has been shown in several previous studies that the significant increase of adhesion and friction is caused by the presence of reactive carbon dangling bonds. However, in humid environments, those dangling bonds can be saturated by hydrogen atoms or hydroxyl groups dissociated from water or hydrogen molecules [17, 19, 36, 37]. Aside from these two mentioned molecules, diamond can also interact with oxygen molecules. Morikawa and colleagues have demonstrated that oxygen molecules could easily react with the C(100) surface and significantly narrow the bandgap of the materials [38]. As illustrated in the study, the reaction with oxygen dramatically alters the morphology of the surface. We expect that the phenomena can also affect the tribological properties of the films. Therefore, knowledge about the reaction mechanism of these molecules with diamond surfaces could open the possibility to better control the friction and wear properties of the films.

Despite the growing interest in understanding the reaction mechanisms, there is still only a very limited amount of work investigating the reaction of the mentioned compounds on different diamond surfaces. The interaction of water and hydrogen molecules with diamond (001) has been previously studied by first-principles calculations [36, 39, 40]. G. Levita et al. showed that the dissociation of water and hydrogen on a diamond C(111)-Pandey surface requires high energy barriers (1.51 and 2.33 eV, respectively), suggesting that the dissociative reactions occur at low kinetic rates [41]. In addition, the chemical reaction of oxygen with diamond (100) surface starting from the adsorption of gas-phase $O_2$ up to the desorption of CO and etching of the surface has been studied by Morikawa et al. [38]. To our knowledge, however, there is still no comprehensive study that compares the adsorption and dissociation of the three molecules on the most stable diamond surfaces. In addition, there is a lack of knowledge in the reaction mechanism of these molecules on the diamond C(110) surface.

Here, we apply first-principles calculations based on density functional theory (DFT) to investigate the adsorption and dissociation mechanism of $H_2$, $H_2O$, and $O_2$ on diamond surfaces. We compare the result on the three types of diamond surfaces, including the C(111)-Pandey-reconstructed, the (2x1)-dimer-reconstructed C(001), and the C(110) surfaces. All these surfaces present a dimer reconstruction composed of carbon double bonds that may resemble the thin DLC layer of $sp^2$ carbon often detected in tribological contacts [42]. By means of first-principles calculations, we aim at determining the adsorption sites and energies for both molecular adsorption and dissociative chemisorption on the three mentioned surfaces. Finally, we determine the barrier energy for the dissociation of $H_2$, $H_2O$, and $O_2$ molecules on the considered surfaces by means of the Nudged Elastic Band (NEB) method [43, 44].

2. **Computational details**

In this work, DFT simulations have been performed using the plane-wave software Quantum ESPRESSO [45]. The exchange-correlation term is described by the Generalized Gradient Approximation (GGA) parametrized by Perdew-Burke-Ernzerhof [46, 47]. As the physisorption of $H_2$ and $H_2O$ on diamond surfaces is dominated by the long-range van der Waals interactions (vdw), a semi-empirical correction proposed by Grimme (D2) [48, 49] was included for all simulations. Due to the large size of our systems (up to 192 atoms) and the high computational demand of the NEB calculations, it is unfeasible to use hybrid functionals. However, we validated the GGA-PBE plus D2 approach that we used by a comparison test with the PBESol functional, which has shown to provide more accurate structural and energetic properties of the material than the conventional PBE functional [50, 51]. The results, which are reported in the SI, show that although the PBESol functional underestimates the adsorption energy, the trend does not change. For example, the order of the molecular adsorption energy ($O_2 > H_2O > H_2$) and dissociative adsorption energy ($O_2 > H_2 > H_2O$) are preserved. Previous studies indicated that the combination of GGA-PBE and D2 could provide the best compromise for accuracy and computational cost [38]. Therefore, the functional has been adapted by a number of publications to study the diamond-related materials [52-54]. We used a convergence threshold of $10^{-4}$ Ry for the total energy and $10^{-3}$ Ry/Bohr for ionic forces. The convergence criteria for the self-consistent electronic (SCF) loop was set at $10^{-6}$ Ry. Since the presence of surface dangling bonds and dissociated molecules could cause possible magnetization in the system, spin polarization is included in all our calculations.

We considered three different diamond surfaces, including the diamond C(110), diamond C(111)-Pandey, and diamond C(001) slabs. A large orthorhombic supercell with minimal lateral dimensions of 8.74 Å is used in all simulations. The large supercell allows us to avoid the interaction of the molecules with their periodic replicas. In particular, the C(001) surface and C(110) were simulated using a 4×4 in-plane supercell, while the C(111) was modeled by a slab in-plane size of 4×3. It is also worth mentioning that the C(001) surface and C(111) slab represent the (2×1) reconstruction. On the other hand, the C(110) is unreconstructed. The thickness of the C(110) slabs was set at 7 atomic. The number of the atomic layer is increased into 8 and 10 layers for the C(111) and C(001), respectively. The chosen thicknesses for the diamond surface models used in this study are based on a number of previous studies [39, 52, 53, 55-57], which indicates the thickness of the slabs is sufficient to produce accurate structural and energetic properties. A vacuum region of 20 Å thickness is included in the supercell to separate each slab from its periodic replicas along the [001] direction. A Monkhorst–Pack (MP) k-point mesh of 2×2×1 and a plane wave cut-off of 30 Ry were applied in all of our simulations. The k-point mesh and cut-off energy have been tested to make sure the error in energy is less than 3 meV/atom.

The adsorption energies, $E_{ad}$, are calculated as $E_{ad} = E_{total} - E_{surf} - N \times E_{mol}$, where $E_{total}$ is the energy of the adsorbate system, $N$ is the number of molecules in the supercell, and $E_{surf}$ and $E_{mol}$ are the energies of the optimized clean surface and the isolated molecule, respectively. Reaction paths for the transition between molecule adsorption and dissociative chemisorption were obtained with the nudged elastic band (NEB) method within the climbing image scheme [43, 44].

## 3. Results and Discussions

### 3.1. $H_2$, $H_2O$, and $O_2$ adsorption on diamond surfaces

In this section, we present the results obtained for molecular adsorption on the clean, reconstructed diamond C(111), C(100), and C(110) surfaces. In order to identify the most favorable configurations, the adsorption energy was calculated for different positions and orientations of the molecules on the surfaces. The results of this analysis, based on the comparison of 44 configurations, are presented as supplemental materials (SI) [Fig. S1-S3], while **Fig. 1** shows the most stable structures that have been identified for each surface. The calculated bond length for the top atomic layer [1.44 Å for C(111), 1.38 Å for C(001), and 1.44 Å for C(110)] of the diamond slabs is shorter than the carbon bond length in bulk diamond (1.55 Å). The result indicates the existence of C=C double bonds of carbon dimmers on the top layer that can be polarized and attract the molecules by vdw interactions. Our calculation indicates that $H_2$ and water molecules only physically adsorb on the three surfaces. In particular, the adsorption energy of $H_2$ molecules on the three surfaces ranges from 0.07 eV to 0.26 eV. Meanwhile, the values obtained for $H_2O$ molecules span from 0.23 eV to 0.67 eV. Our data is in good agreement with previous calculations [36, 39-41]. In both cases, the adsorbate tends to lay on top of the surface trenches with an equilibrium distance from the nearest carbon atom of the diamond surfaces greater than 2.28 Å. The stronger adsorption of the $H_2O$ molecules could be due to the higher polarization of the water molecule. As a result, water interacts stronger with these diamond surfaces by pointing its partially charged H atoms towards C atoms partially charged with opposite sign (**Fig. 1a-f**) [39]. Noticeably, the adsorption strength increases significantly in the case of C(110), as shown in **Fig. 1a-f** and **Fig. 2a**. While the adsorption energy of $H_2$ and $H_2O$ is relatively similar for the C(111) and C(001) surfaces, those values for C(110) increased by threefolds. The results suggest that the C(110) surface is more chemically active compared with the C(111) and C(001) ones. Interestingly, we found that the adsorption energy increases as the surface energy increase. Previous studies showed that the surface energy increased from 4.06 $J/m^2$ to 5.71 $J/m^2$ and 5.93 $J/m^2$ for the C(111), C(001), and C(110), respectively [58]. This order is the same for the adsorption energy, as can be seen from **Fig. 2a**.

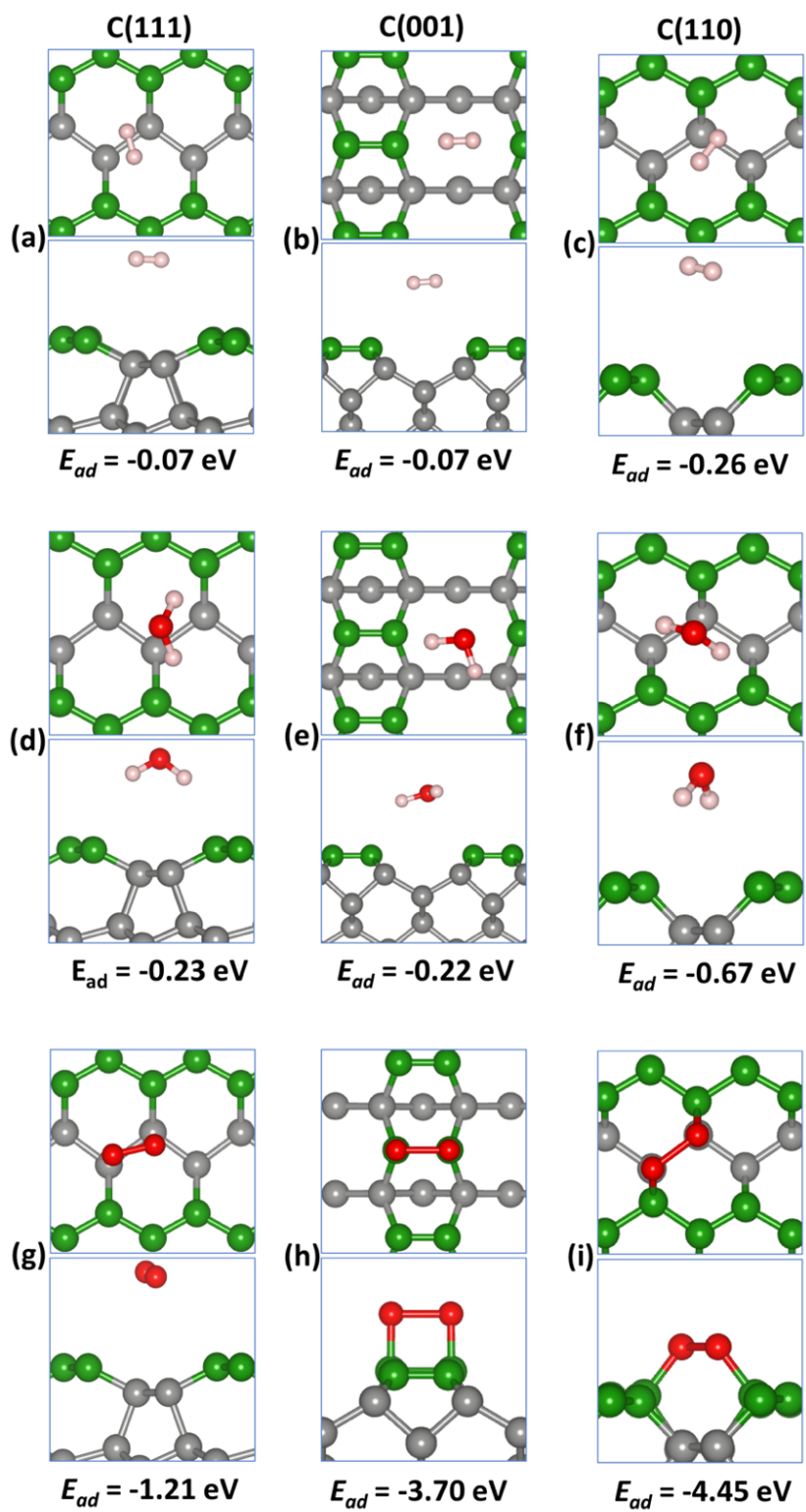

***Fig. 1.*** *Most stable adsorption geometries of $H_2$, $H_2O$, and $O_2$ molecules with corresponding adsorption energies on three diamond surfaces [top view (upper) and side view (lower) panels]. C on the top layer (Green), C (gray), O (Red), H (White). The color code is applied throughout this work.*

The adsorption of oxygen molecules on the three diamond surfaces is much more favorable compared to $H_2$ and $H_2O$, with a mix of physical adsorption and chemical adsorption. In the case of C(111), the adsorption distance is still in the range of physical adsorption, with the adsorbate laying on top of the surface trenches. However, the adsorption energy (1.21 eV) is much larger compared to that of $H_2$ and $H_2O$. The $O_2$ molecule can chemically adsorb on the C(001) surface with a larger adsorption energy of 3.70 eV, indicating that this configuration is very stable. As shown in **Fig 1h**, the $O_2$ molecule adsorbs on top of the carbon dimer. The calculated bond lengths indicate that both the O=O bond of $O_2$ and C=C bond of carbon dimers below are stretched from 1.23 Å and 1.44 Å to 1.52 Å and 1.54 Å, respectively. The obtained values are close to the bond length of O-O (1.48 Å – 1.5 Å) [59] and C-C (1.55 Å in the bulk diamond) single bonds indicating that the adsorption can change the double bond of both the oxygen molecule and the carbon dimers below it into single bonds. Interestingly, the chemical adsorption of $O_2$ on C(110) is not on top of the C=C bonds (**Fig. 1i**). Instead, the molecule lays above the surface trench, with each oxygen atom bonded two carbon atoms from different carbon chains. The energy gain for adsorption (4.45 eV), in this case, is even higher than that of the C(001) surface, indicating a strong chemical activity of the C(110) surfaces. Although there is a mix of physical and chemical adsorption, the adsorption energy of the $O_2$ molecule follows the same trend observed for $H_2$ and $H_2O$, according to which the adsorption energy increases with the surface energy.

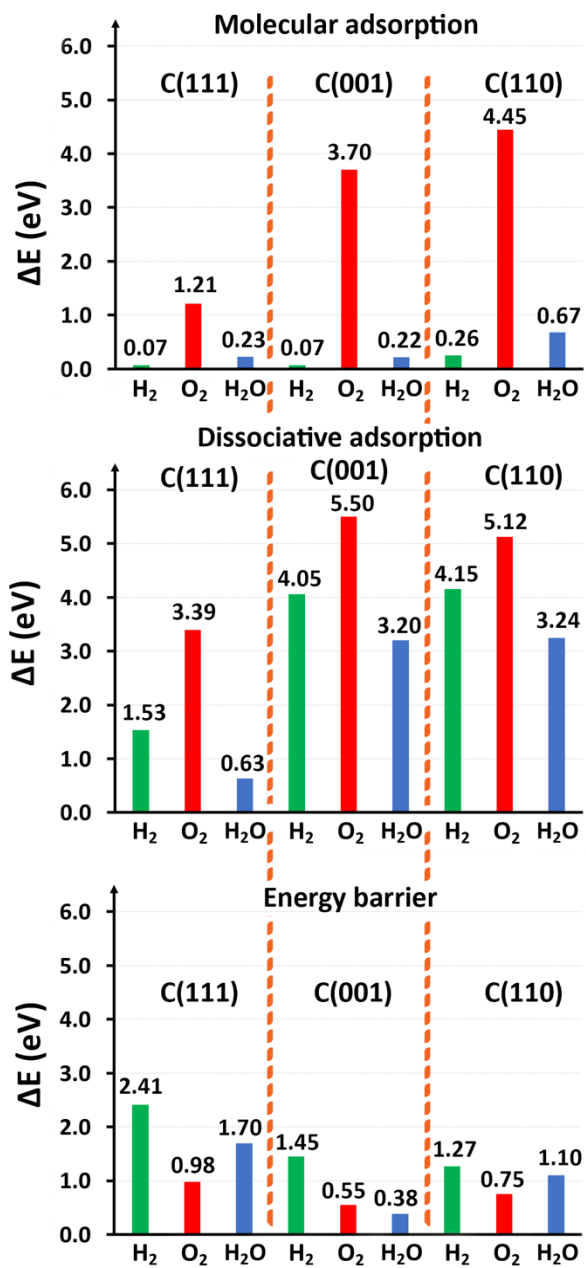

*Fig. 2.* Comparison of the energy for adsorption, dissociative chemisorption, and dissociation barrier for $H_2$, $H_2O$, and $O_2$ on the three considered diamond surfaces.

### 3.2. Dissociative adsorption of $H_2$, $H_2O$, and $O_2$ on diamond surfaces

Next, we investigate the dissociative adsorption of the three molecules on diamond surfaces. **Fig. 3** illustrates the most stable configurations, while all the 46 considered configurations are reported as SI (**Fig. S4-S6**). The energy gain for dissociative chemisorption is significantly higher than that obtained for molecular adsorption for all the considered molecules (**Fig. 2b**).

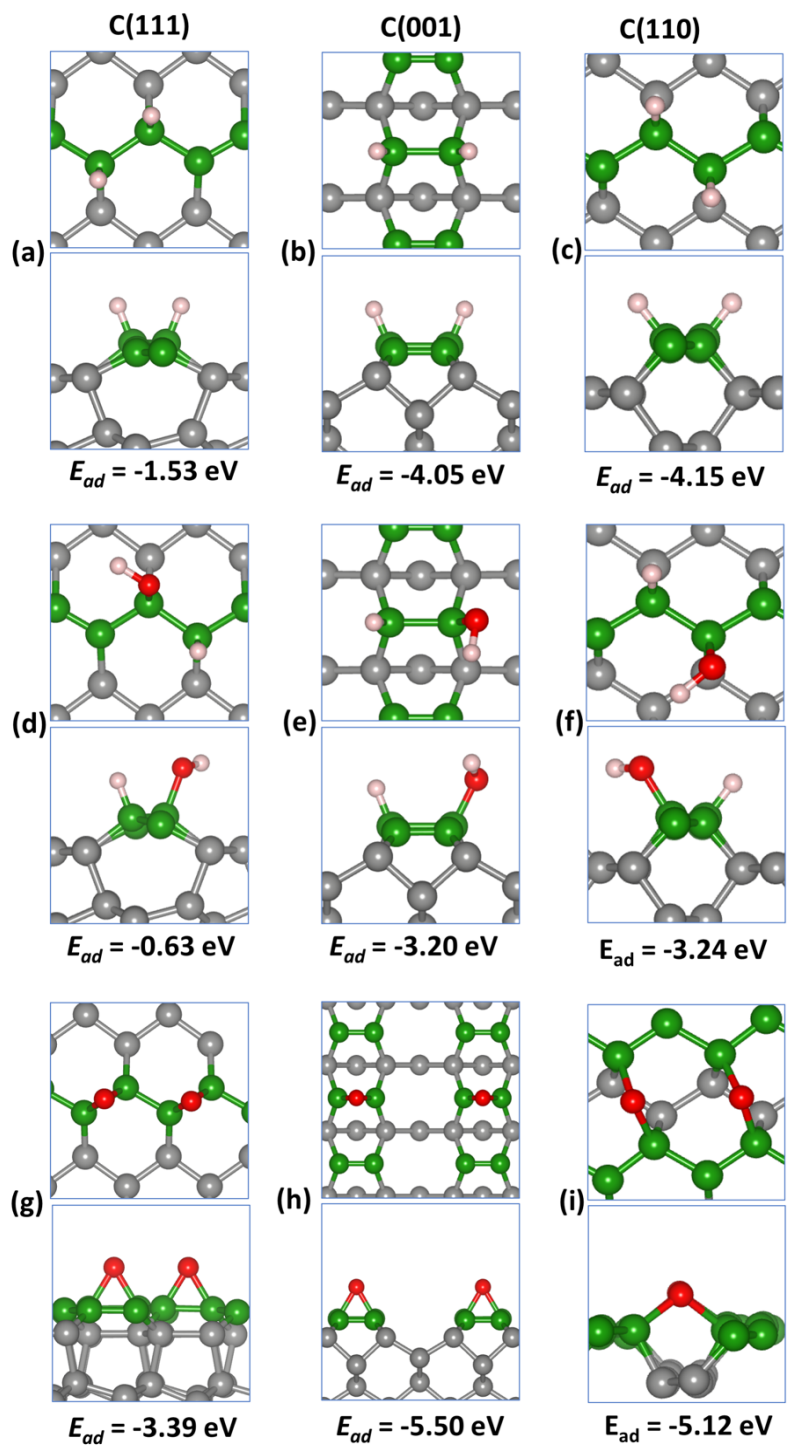

*Fig. 3*. *Dissociative adsorption configurations of H₂, H₂O, and O₂ molecules with corresponding adsorption energies on three different types of diamond surface [top view (upper) and side view (lower) panels].*

Particularly, H atoms prefer to adsorb on top of the C dimer, with adsorption energy of 1.53 eV, 4.05 eV, and 4.15 eV for C(111), C(001), and C(110), respectively. Our data is in good agreement with the previous calculation of 1.61 eV and 3.7-4.3 eV for C(111) [41] and C(001) [60-62] surfaces, respectively. Due to the electrostatic (repulsive) interaction between two positive charged atoms [41], the two H atoms point away from each other, as shown in **Fig. 3a-c**. The adsorption breaks the C=C double bond of the C dimer below. The calculated bond length indicates that the C-C bond of the C dimer increased from 1.44 Å, 1.38 Å, and 1.44 Å to 1.60 Å, 1.61 Å, and 1.55 Å for C(111), C(001), and C(110), respectively. A similar trend is observed in the case of the water molecule. While the energy gain for dissociative chemisorption for $H_2$ and $H_2O$ on diamond surfaces increases as C(111) > C(001) > C(110), which is similar to the case of molecule adsorption, the tendency is broken for the case of the $O_2$ molecule. As shown in **Fig. 2**, the energy gain for $O_2$ dissociation on C(001) is slightly larger than C(110). Interestingly, the energy gain for $H_2$ dissociation (from 1.53 eV to 4.15 eV) is higher than $H_2O$ (from 0.63 eV to 3.24 eV) in all three diamond surfaces. The strong adsorption of $H_2$ could be beneficial for tribological applications as a previous study indicated that the H-passivated diamond C(001) surface provides better fiction reduction compared with OH-passivated one [60]. It is worth mentioning that the adsorption of both hydrogen and hydroxyl groups could help to lower the friction of diamond surfaces under humid environments. Numerous experimental and theoretical studies have clearly highlighted the crucial roles of adsorbed hydrogen and hydroxyl groups on the hydrophilic, electronic, and mechanical behavior of diamond-like materials. The presence of reactive dangling bonds on the clean DLC surfaces could significantly increase adhesion and friction. Still, those dangling bonds can be saturated by hydrogen and hydroxyl groups dissociated from water or hydrogen molecules [17, 19, 36, 37, 63]. The present results indicate that once formed, the C-H bonds are expected to resist subsequent dissociation due to, e.g., the wear-off caused by rubbing against a countersurface, more efficiently than the C-OH bonds.

The dissociative adsorption of $O_2$ on the three diamond surfaces is quite different compared to $H_2$ and $H_2O$, which is due to the ability of O atoms to form two covalent bonds. As shown in **Fig. 3g-h** for the C(111) and C(001) surfaces, the O atoms prefer to adsorb on the bridge of the C-C bonds rather than on top of the C atoms, as it happens in the case of $H_2$ and $H_2O$. Each O atom forms two covalent bonds with the underneath C atoms. The configuration is expected to break the C=C double bond of the below C dimmer in a similar mechanism as $H_2$ and $H_2O$. Our calculated bond length indicates that C-C bonds increase from 1.38 Å - 1.44 Å to 1.50 Å. Meanwhile, on the C(110) surface, we found that the O atom lay on top of the surface trenches, forming two covalent bonds with C atoms from two different chains, as shown in **Fig. 3i**. We did not observe this configuration for the case of C(111), even though the two diamond slabs own similar surface morphology (zig-zag carbon chains on the top layer). This could be because the distance between two zig-zag carbon chains on C(111) (3.90 Å) is much larger than in C(110) (3.13 Å). Therefore, the O atom could not form a bridge between two chains. Furthermore, while scanning the favorable sites for the O atom on the C(110) surface, we found that the adsorption of the O atom could induce the out-of-plane displacements of C atoms from the surface. Specifically, the C atom could be pulled out from the surface, as shown in **Fig. 4**. The result suggests that the interaction of the molecule with C(110) under tribological conditions such as shearing and high temperature could potentially wear this surface.

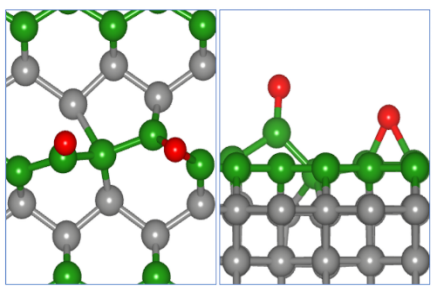

*Fig. 4. Out-of-plane displacement of C atoms induced by Oxygen chemisorption on the C(110) surface, top (lateral) view in the left (right) panel.*

Therefore, in the next sections, we will discuss in detail the reaction mechanism of the three molecules and how the interaction could alter the surface morphology.

### 3.3. Reaction paths for $H_2$ and $H_2O$ dissociative chemisorption on diamond surfaces

To further characterize the interaction between the $H_2$ and $H_2O$ with the three diamond surfaces, we calculate the reaction path and associated activation energy for the dissociation. The NEB method is employed for this purpose. We consider the most stable molecule adsorption configuration (**Fig. 1**) as the initial states for the reaction, and the final states correspond to the most stable dissociative adsorption configurations (**Fig. 3**). This choice is justified by the fact that the most stable configurations have the highest statistical weight. However, one should keep in mind that the barrier height may be affected by the choice of the initial and final states. Our data in Fig. S1-S3 indicates that except for the unfavorable adsorption sites (positive adsorption energy), the reaction energy of $H_2$ and $H_2O$ molecules diffusing between different adsorption positions is in the small range of 0.00-0.05 eV and 0.00-0.49 eV, respectively. Meanwhile, due to the chemical adsorption, the highest reaction energy for the diffusion of the $O_2$ molecule is 3.10 eV on C(110). In addition, the barrier energy is often correlated with the reaction energy [64]. The result indicates that the $H_2$ and $H_2O$ can easily diffuse back and forth between adsorption sites, while it is more difficult in the case of the $O_2$ molecule. Although the diffusion of the molecules and their dissociated atom on different diamond surfaces is a very interesting topic, we did not focus on this part due to its complicated simulations that could make the manuscript come lengthly. This is because there are many different adsorption sites leading to a large number of possible diffusion paths to be considered by NEB calculations.

The outcome of the NEB analysis for the dissociation of $H_2$ on the three diamond surfaces is reported in **Fig. 5**. The energy barrier estimated for the dissociation of $H_2$ on C(111) Pandey is very high (2.41 eV), in good agreement with a previous study (2.33 eV) [41]. The result suggests that it is improbable for the reaction to happen on the reconstructed C(111) Pandey and highlights the relatively inert behavior of the Pandey reconstruction towards hydrogen splitting. However, once $H_2$ dissociates and chemically adsorbs on the surface, it becomes difficult to detach the molecule with a barrier energy of 3.87 eV. The calculated barrier energy is close to what has been reported from experimental work (3.7 eV) using Temperature programmed desorption (TPD) analysis [18]. The energy barrier is strongly reduced in the case of C(001) (1.45 eV). However, it is still relatively large, with the C(110) being the most chemically active surface with a barrier energy of 1.27 eV. As expected, we found that the energy barrier for $H_2$ dissociation follows the opposite energy trend compared with the molecular and dissociative adsorption. In particular, the activation energy is reduced as the surface energy increases C(111) > C(001) > C(110) (**Fig. 2c**).

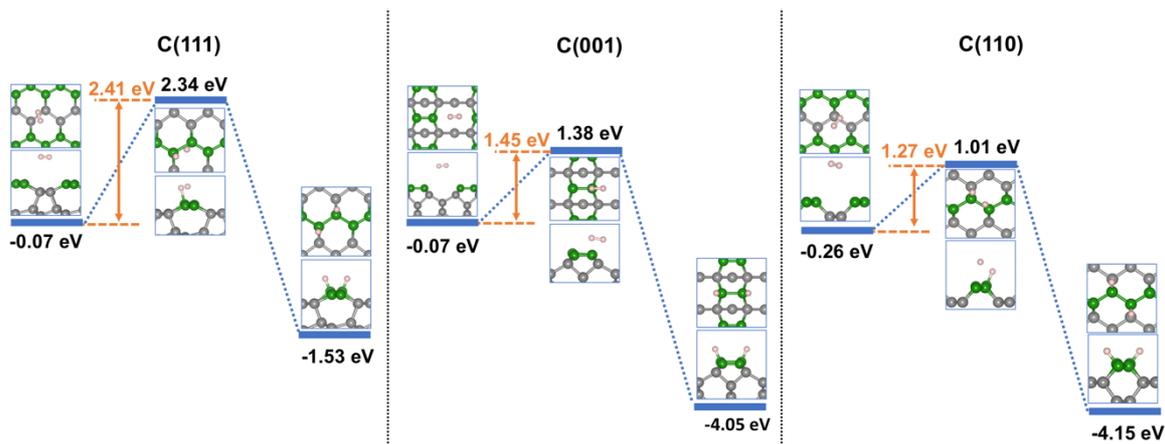

*Fig. 5. Initial, transition, and final states of the reaction paths calculated for H₂ dissociation by means of the NEB method. The total energy of the gas-phase H₂ molecule and the clean diamond surfaces is taken as the reference. The activation energies are noted in orange.*

The calculated activation energy for $H_2O$ shown in **Fig. 6** indicates that the $H_2O$ molecule is easier to dissociate on C(111) and C(001) surfaces compared to $H_2$. Its barrier energy on C(111) Pandey surface is 1.70 eV, while the barrier on C(001) is only 0.38 eV. The lower activation energy could increase the chance that the surface is passivated by hydrogen and hydroxyl groups under tribological conditions. The passivation could reduce friction by preventing the formation of covalent bonds across the opposing sliding surfaces [17, 32, 33]. Furthermore, the adsorbed hydroxyl are expected to enhance the interaction with other water molecules through hydrogen bonding [39, 40, 65]. The mechanism could be a major atomic-scale driving force for the increased hydrophilicity of DLC surfaces. Ab initio calculations have shown that the formation of a wetting layer could further prevent the formation of covalent bonds at interfaces and make them slippery [32, 33, 42], thus reducing friction in humid conditions. The mechanism is based on the fact that the dangling bond carbon atoms on some diamond surfaces are highly chemically active and could form covalent bonds across interfaces, thus increasing the friction. However, we found that the C(111) Pandey reconstructed surface is very inert to chemical reactions. Therefore, it might not be necessary to passivate this surface to reduce the friction. Specific computational studies that take into account the chemical composition and structure of the countersurface are necessary to answer this question.

Interestingly, while the NEB analysis showed that C(111) Pandey is the most inert diamond surface for the dissociation of both $H_2$ (2.41 eV) and $H_2O$ (1.70 eV) molecules, there is a striking difference for the C(001) and C(110) surfaces. In particular, the barrier energy for the reaction of $H_2O$ on C(110) (1.10 eV) is much higher than that of C(001) (0.38 eV), which is opposite to the case of $H_2$. The result suggests that the barrier energy for the dissociation of $H_2O$ may also depend on the morphology of the surfaces. While the top layer of C(111) and C(110) is characterized by the zig-zag carbon chains with different distances between each chain, the C(001) features separated C=C dimers. In addition, the shape of the $H_2O$ molecule structure could make it difficult for the molecule to rotate and react on the surface compared to that of $H_2$ and $O_2$. We explain this behavior by inspecting the transition state in the three diamond slabs. We find that in both C(111) and C(001) cases, the water molecules first form a covalent bond between O with a carbon atom below. This configuration could facilitate the dissociation of the water molecule since it weakens the O-H bonds as the O atom has to share its electrons to form the new C-O covalent bond. However, in the case of C(110), the stronger adsorption of the molecule make it difficult for the molecule to rotate the O atom toward the surface. As a result, the molecule follows a different mechanism and donates the H atom to the surface first, and then the hydroxyl group will adsorb the latter. This new reaction mechanism explains why the barrier energy of $H_2O$ does not follow the energy trend as in the case of $H_2$. It is worth mentioning that the low barrier energy for the dissociation of water on C(001), indicating the interaction of the molecule with the surface will mainly produce C-H and C-OH bonds. Our result is in good agreement with the high-resolution electron-energy loss spectroscopy (HREELS) measurements by Gao et al., [66]. However, under the tribological condition, the mechanical energy supply can facilitate higher activation energy reactions. Indeed the spectroscopic analysis performed by Konicek et al. [19] revealed an increase of C-O and C = O bondings after the tribological test.

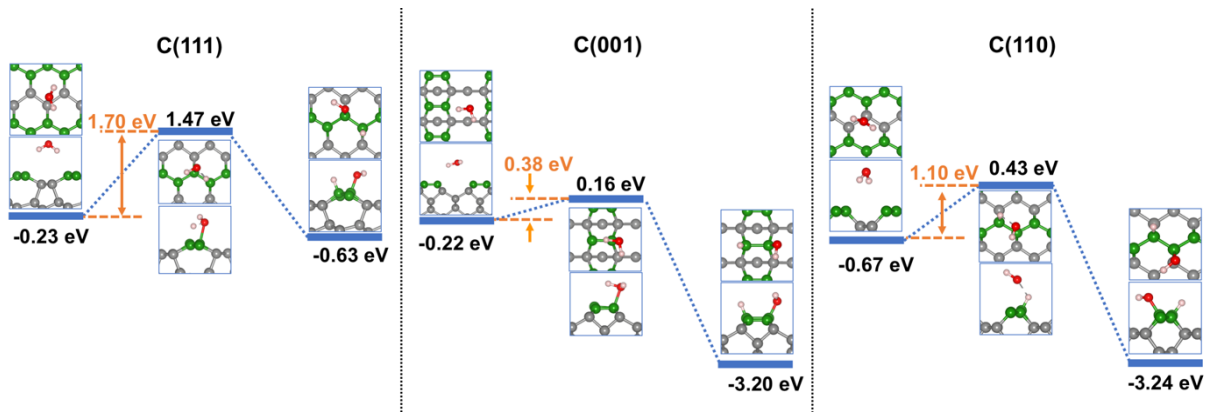

*Fig. 6*. Initial, transition, and final states of the reaction paths calculated for H$_2$O dissociation by means of the NEB method. The total energy of the gas-phase H$_2$O molecule and the clean diamond surfaces is taken as the reference. The activation energies are noted in orange.

### 3.4. Reaction mechanism of O$_2$ on diamond surfaces

Next, we investigated the dissociation mechanism of O$_2$ on the three diamond slabs. As mentioned previously, when scanning the favorable sites for the adsorption of O atoms on the C(110) surface, we found that the adsorption of the O atom could induce the out-of-plane displacement of carbon atoms. In addition, while performing the NEB analysis using the same procedure as for H$_2$ and H$_2$O molecules, i.e., using the most stable molecule adsorption configuration as initial state and dissociative geometry as the final state, we found that the reaction path could pass through several intermediates. These reaction steps could significantly alter the structure of the diamond surfaces. In fact, the oxygen-induced deconstruction of the C(100) surface has been reported in electron diffraction experiments [67, 68]. Also, a previous study indicated that the O atom in silica could form a strong bond with C(110) surface initiating the wear of C atoms from the diamond surfaces [69]. Understanding those reaction mechanisms could open the possibility to reduce the wear of DLC coating layers. Therefore, in this section, we not only discuss the dissociation of the O$_2$ molecule but also how they could induce the deconstruction of the diamond surfaces.

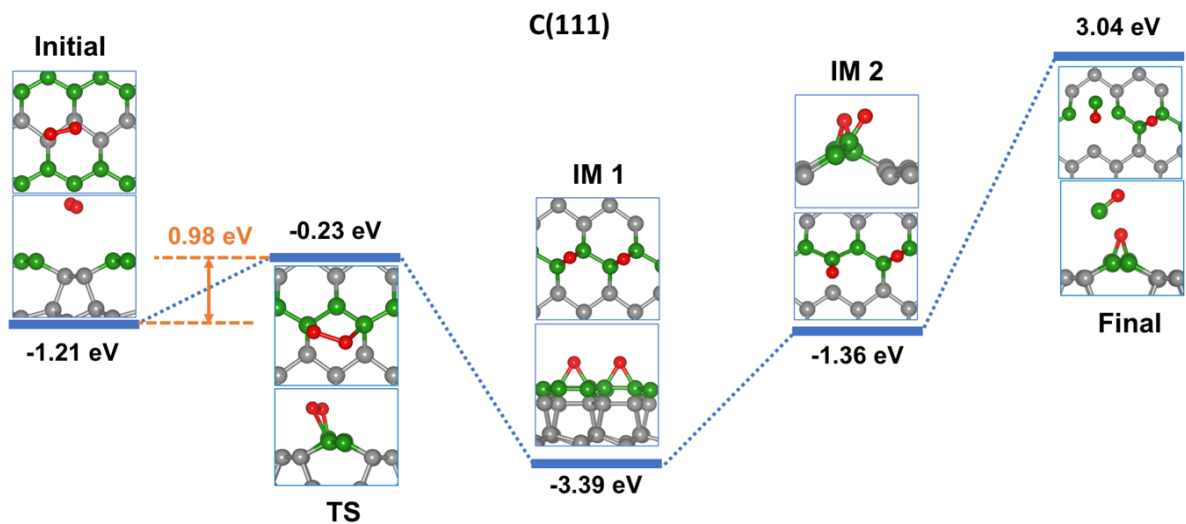

***Fig. 7.*** *Oxygen-induced surface modification after $O_2$ adsorption and dissociation on the diamond C(111) surface according to NEB calculations. The plot uses the total energy of the gas-phase $O_2$ molecule and the clean C(111) surface as the reference. The activation energies are noted in orange.*

**Fig. 7** shows the reaction path of $O_2$ from molecule adsorption on the C(111) surface. It is clear that $O_2$ is more chemically active than $H_2$ and $H_2O$, indicated by a lower barrier of 0.98 eV for the dissociation compared with 2.41 eV and 1.70 eV for the case of $H_2$ and $H_2O$, respectively. Starting from the dissociative adsorption configuration, we also investigate the possibility for the carbon atom to be pulled out and completely detached from the C(111) surface through the desorption of the CO molecule. The result in **Fig. 7** indicates extremely high barrier energy of 2.03 eV and 4.40 eV for pulling out the surface carbon atoms and desorbing them in the form of CO. These high barrier energies make the reactions very unlikely. In addition, there is no barrier energy for the opposite reaction. Therefore, the CO molecule could easily re-adsorb to the surface. As a result, our calculation suggested that it is very difficult for the C(111) reconstructed surface to be deformed under the reaction with the $O_2$ molecule.

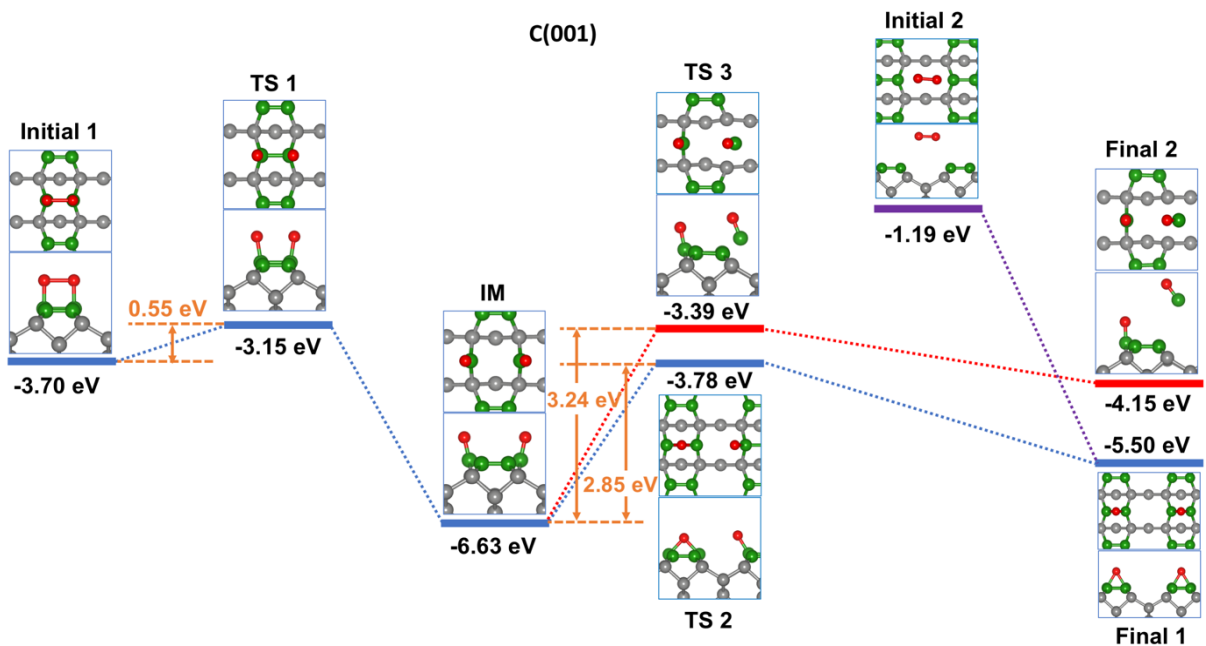

**Fig. 8**. Oxygen-induced surface modification after $O_2$ adsorption and dissociation on the diamond C(001) surface according to NEB calculations. The plot uses the total energy of the gas-phase $O_2$ molecule and the clean C(001) surface as the reference. The activation energies are noted in orange.

The dissociation of $O_2$ on the C(001) reconstructed surface is much more complicated compared with the case of C(111). Our NEB analysis indicates that in order to dissociate the molecules starting from the molecule adsorption to dissociative adsorption, the reaction needs to undergo an intermediate. As shown in **Fig. 8**, after chemically adsorbing on the surface, the $O_2$ molecule is dissociated together with the C-C bond below to form the Ketone configuration with a small barrier energy of 0.55 eV. Compared to the case of C(111), the dissociation of the $O_2$ molecule on C(001) alters the structure of the surface as the C-C bond is broken. We found that the Ketone configuration is very stable with an adsorption energy of 6.63 eV. The barrier energy for the opposite reaction is enormous, indicating it is extremely difficult for the $O_2$ molecule to be reformed and desorb from the surface. Starting from the Ketone state, one of the O atoms needed to transfer on top of the bridge site to form the dissociative configuration, as shown in **Fig. 8** (Final 1). The high barrier energy of 2.85 eV indicates this is the rate-limiting step for the reaction path.

Since it is hard to form the dissociative adsorption configuration with the molecule adsorption as the initial state, we have tested the possibility where the $O_2$ molecule directly dissociates from the physical adsorption as illustrated by the purple line in **Fig. 8**. We found that the activation energy turned out to be negligible (0.02 eV), indicating that the dissociative adsorption of the $O_2$ molecule could potentially cause by the direct dissociation of $O_2$ when the molecule is floating on the surface. As long as the $O_2$ molecule chemically adsorbs on the surface, it will be more favorable to form the Ketone configuration with higher adsorption energy. It is worth mentioning that, in this work, we did not consider the transition from Ketone to Ether configuration. We found that the formation of Ether configuration is less favorable compared to Ketone under the low coverage condition. This result is consistent with the previous report by Morikawa et al., indicating that the state can only be formed at high coverage adsorption of O atoms [38]. Therefore, the most promising path is to initiate the Ketone configuration. After the coverage of the Ketone configuration reaches a certain threshold, it can facilitate the formation of the Ether configuration. Our result is in good agreement with experimental data by Hossain et al. [70]. In their study, surface vibration probes using Electron energy loss spectra (EELS) indicate a loss peak at 215 meV when the surface is exposed to a low exposure of oxygen, which has been assigned to carbonyl stretching (Ketone). As the oxygen exposure increased (30L), the loss peaks of the pure C(001) surface were replaced by loss peaks at 113 meV and 150 meV indicating the carbonyl bending and ether stretching, respectively. Similarly, John et al. [71] used high-resolution x-ray photoelectron spectroscopy (XPS) to demonstrate that the C(100) surface could form carbonyl and ether groups with coverage of up to a monolayer.

We calculate the desorption of CO from the surface by considering Ketone as the initial configuration for our NEB analysis, as shown by the red line in **Fig. 8**. We chose this initial configuration since it is more reasonable for the $O_2$ molecule to dissociate and break the C=C double before the detaching of the CO molecule. As shown in **Fig. 7**, the barrier energy for the reaction is 3.24 eV which is much lower than that of the C(111) surface. However, this barrier is still very large, indicating the reaction could only happen at elevated temperatures or under high shear stress (under the tribological condition).

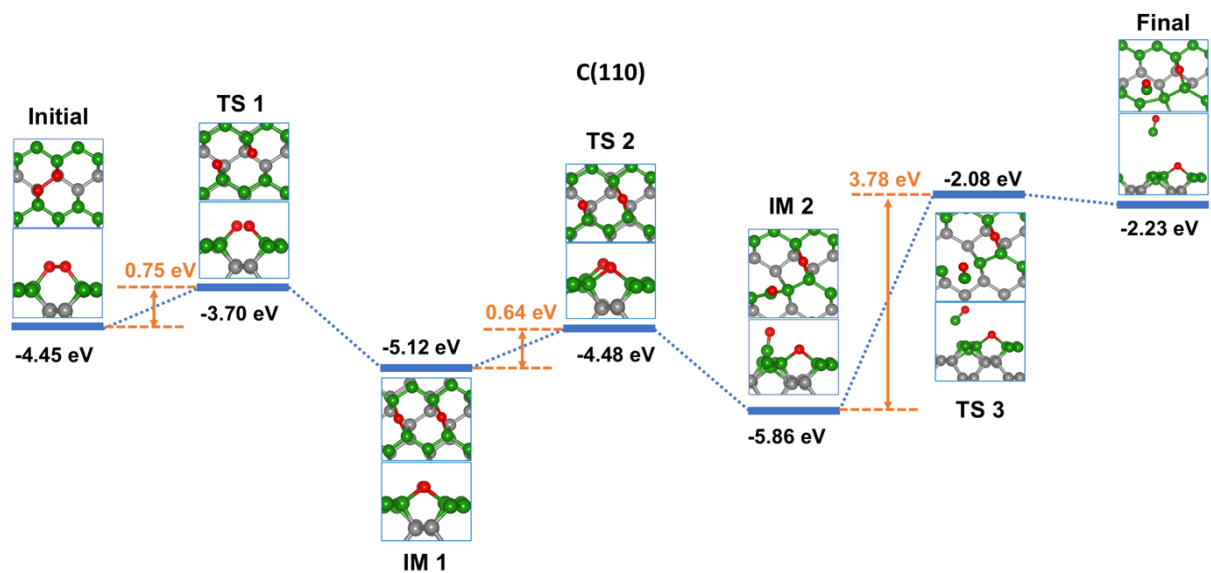

**Fig. 9**. Oxygen-induced surface modification after $O_2$ adsorption and dissociation on the diamond C(110) surface according to NEB calculations. The plot uses the total energy of the gas-phase $O_2$ molecule and the clean C(110) surface as the reference. The activation energies are noted in orange.

Similarly, we also perform NEB analysis for the reaction of $O_2$ molecules on the C(110) surface by considering the most stable chemical (molecule) adsorption as the initial state (**Fig. 9**). Different from the case of the C(001) surface, the $O_2$ molecule does not undergo an intermediate state to form the dissociative adsorption form. The barrier energy for the reaction is relatively small (0.75 eV) compared to the rate determined step (2.85 eV) as in the case of the C(001) surface. The result indicates the likelihood of forming the dissociation adsorption configuration from the chemical (molecule) adsorption is much higher compared to the case of C(001). In addition, we found that there is no modification in the structure of the diamond surface caused by the reaction.

In this work, we also investigate the possibility of altering the surface by atomic adsorption. One carbonyl group is pulled out from the surface, as shown in **Fig. 9** (IM 2). We found that this configuration is more stable than the dissociative adsorption. In addition, the barrier energy for the transformation is only 0.64 eV. The result suggested that after the dissociation of the oxygen molecule, the diamond surface could undergo a significant change as the carbonyl group is pulled out. A previous study by Chaudhuri et al. [72] has proved the coexistence of carbonyl and ether groups on the C(110) surface by using primarily computational methods and performed experiments to validate their computational results. In addition, Hussein et al. [52] also indicated that a mixture of ether (C−O−C) and ketone (C=O) groups on the C(110) surface was found to be the most energetically favorable configuration. We further study the detachment of the CO molecule from the surface by calculating the barrier energy for the desorption. Similar to the case of the C(001) surface, we found that the desorption of CO molecules from the surface is unlikely, with a high activation energy of 3.74 eV. The result indicates that even the chemical adsorption and dissociation of $O_2$ molecule could induce the deformation of the C(001) and C(110) slabs, it is still difficult for the carbon atom to be worn out of the surface in the form of CO molecule. However, the detachment of the C atom caused by the contemporary chemisorption into a counter surface may result in more favorable and cause the wear of diamond. The deconstruction of the diamond surface would require extreme tribological working conditions such as high temperatures or high shear stress. It is worth mentioning that the formation of the carbonyl groups on the diamond C(001) and C(110) surfaces could significantly alter the tribological properties of the materials. On the one hand, the carbonyl group could potentially increase the friction and wear due to the higher chemical activity of the oxygen in the group compared to H atoms, which is due to C=O electronegativity attributed to the oxygen and its two lone pairs of electrons. On the other hand, the surface carbonyl O atoms could enhance hydrophilicity by interacting with subsequence water molecules in the air. The two lone pairs of electrons make the oxygen more electronegative than carbon. Therefore, the group can attract water molecules through hydrogen interaction. The process could form a water layer separating the two sliding surfaces from close contact and reducing friction.

## 4. Conclusions

In this study, the physical and chemical adsorption of $H_2$, $H_2O$, and $O_2$ molecules on the most stable diamond surfaces, including C(111), C(001), and C(110), are investigated by means of first-principles calculations. The most stable phases for the C(111) and C(001) surfaces were considered, namely the Pandey and the (2×1)-dimer reconstruction, respectively. Therefore, all the three considered surfaces terminate with a layer of C=C bonds. The most favorable configurations for molecular adsorption and dissociative chemisorption are identified for the three surfaces by comparing several possible configurations. The reaction paths for the transition between the most stable molecular- and dissociative-adsorption configurations are identified. In addition, the deconstruction mechanism of these diamond surfaces caused by the reaction with oxygen was also examined. Our result can be summarized as follow.

- There is a general trend in the molecular adsorption energy of $H_2$ and $H_2O$ on the three diamond surfaces. In which the energy gain for adsorption increases as the surface energy increases with the order C(111) < C(001) < C(110). This tendency is broken in the case of oxygen dissociative adsorption. Nonetheless, the C(111) Pandey-reconstructed surface is still the most chemically inert surface indicated by the lowest adsorption energy in both molecular and dissociative adsorption forms. In addition, oxygen is the most chemical active compound with significantly higher adsorption energy and the ability to undergo both molecular physisorption and chemisorption.
- The dissociative chemisorption is energetically favorable for all the molecules on all the three considered surfaces. Upon adsorption of the molecular fragments, the C=C double bonds are broken to form covalent bonds with the adjacent C atoms on the same upper chain or dimer.
- Despite the fact that dissociative chemisorptions of hydrogen are on the three diamond surfaces more exothermic than that of water, the process requires much higher activation energy. The slow reaction rates suggest that the diamond surfaces are likely to be passivated by both oxygen and dissociated water molecules in humid environments.
- The dissociation processes of oxygen molecules on these diamond slabs are much more complicated than the case of $H_2$ and $H_2O$ with multiple intermediates states. These reactions alter the surface morphology of the diamond slabs, which could significantly affect the tribological properties of the diamond-based coatings. We found that the complete detachment (wear) of the C atom from the slab in the form of a CO molecule has a high activation energy. Nonetheless, the wear of the diamond surface could be easier after oxidation and in the presence of a countersurface.

**Acknowledgments**

These results are part of the "Advancing Solid Interface and Lubricants by First Principles Material Design (SLIDE)" project that has received funding from the European Research Council (ERC) under the European Union's Horizon 2020 research and innovation program (Grant agreement No. 865633). PRACE is also acknowledged for computational resources.